\documentclass[fleqn,usenatbib]{mnras}

\usepackage{newtxtext,newtxmath}

\usepackage[T1]{fontenc}
\usepackage{orcidlink}
\DeclareRobustCommand{\VAN}[3]{#2}
\let\VANthebibliography\thebibliography
\def\thebibliography{\DeclareRobustCommand{\VAN}[3]{##3}\VANthebibliography}

\usepackage{graphicx}	
\usepackage{amsmath}	

\title[On DESI's DR2 exclusion of $\Lambda$CDM]
{On DESI's DR2 exclusion of $\Lambda$CDM}

\author[Marina Cort\^{e}s and Andrew R.~Liddle]{Marina Cort\^{e}s \orcidlink{0000-0003-0485-3767} and Andrew R.~Liddle \orcidlink{0000-0002-8164-4830}\\ 
Institute of Astrophysics and Space Sciences, Faculty of Sciences, University of Lisbon, 1769-016 Lisbon, Portugal}

\date{\today}

\pubyear{2025}

\begin{document}
\label{firstpage}
\pagerange{\pageref{firstpage}--\pageref{lastpage}}
\maketitle

\begin{abstract}
The DESI collaboration, combining their Baryon Acoustic Oscillation (BAO) data with cosmic microwave background (CMB) anisotropy and supernovae data, have found significant indication against the $\Lambda$CDM cosmology. This can also be interpreted as the significance of the detection of the $w_a$ parameter that measures variation of the dark energy equation of state. DESI's DR2 article quotes exclusion of $\Lambda$CDM for combinations of BAO and CMB data with each of three different and overlapping supernovae compilations (at 2.8-sigma for Pantheon+, 3.8-sigma for Union3, and 4.2-sigma for DESY5). We show that one can neither choose amongst nor average over these three different significances. We demonstrate how a principled statistical combination yields a combined exclusion significance of 3.1-sigma. Further we argue that, faced with these competing significances, the most secure inference from the DESI DR2 results is the 3.1-sigma level exclusion of $\Lambda$CDM obtained from combining DESI+CMB alone, omitting supernovae.
\end{abstract}




%
\section{Introduction}

Considerable excitement and debate has been caused by the Dark Energy Spectroscopic Instrument (DESI)'s finding of strong hints of evolving dark energy \citep{DESIVI}, when combining their own baryon acoustic oscillation (BAO) data with cosmic microwave background (CMB) data from the {\it Planck} satellite \citep{Planck,Planck2} and the Atacama Cosmology Telescope (ACT, \citealt{ACT}). The DESI second data release (DR2) has added some further strength to the statistical significance \citep{DESIDR2,Lodha}.

The highest levels of significance are obtained when the BAO and CMB datasets are combined with compilations of supernovae (SNe) distance--redshift relations. Three separate but partially overlapping SNe compilations are available; Pantheon+ \citep{PantheonPlus,Brout22}, Union3 \citep{Union3}, and Dark Energy Survey Year 5 (DESY5, \citealt{DESY5,DESY5data}). Combined with DESI BAO and CMB, they yield exclusions of $\Lambda$CDM versus a simple evolving dark energy model at 2.8-sigma, 3.8-sigma, and 4.2-sigma respectively. Our paper addresses the interpretation of the overall exclusion level on $\Lambda$CDM taking all these results into account.

We focus on the results presented for the CPL (Chevallier--Polarski--Linder) dark energy parametrisation used in the DESI DR2 key cosmological constraints paper \citep{DESIDR2}. The CPL model, also known as the w0waCDM model, is a phenomenological model of dark energy featuring a two-parameter dark energy equation of state \citep{CPL1,CPL2}
\begin{equation}
    w(a) = w_0 + w_a(1-a) \,,
\end{equation}
where $a$ is the scale factor normalized to unity at present, and $w_0$ and $w_a$ are constants. In \cite{DESIDR2} their priors are taken to be uniform in the ranges $[-3,1]$ and $[-3,2]$ respectively, with the additional condition $w_0 + w_a <0$ to allow early matter domination. The cosmological constant, or $\Lambda$CDM, model corresponds to \mbox{$w_0=-1$} and $w_a = 0$, and a regime in which $w(a)<-1$ is called a phantom regime \citep{Caldwell}. Transition across $w=-1$ is referred to as phantom crossing.

The DESI collaboration have additionally carried out further more detailed investigation of dark energy implications \citep{CalderonDR1,LodhaDR1,Lodha,Gan}, primarily in support of the key paper result. There is already a massive literature following the very similar DESI DR1 result \citep{DESIVI}. This literature contains a wide range of views, both supportive and critical, most of which continue to apply to DR2. The interpretation of the results, therefore, remains very much an open issue.

All the collaborations discussed here have put outstanding efforts into data acquisition and analysis, providing the cosmology community with remarkable resources. We take these entire processes, leading up to the construction of data likelihood functions, at face value, and address only issues of the final interpretation of the results obtained.

\section{The nature of the exclusion}

Constraints on $w_0$ and $w_a$ are strongly correlated, as there are no data directly constraining the present-day value of $w_0$. This correlation can be exactly removed by the freedom to choose a pivot redshift \citep{HT,DETF,LinderPivot,MartinPivot} to specify the amplitude, $w_{\rm pivot}$, as we described for DESI DR1 in \cite{CLDESI}. The pivot redshift indicates where the data are most constraining. In a gaussian approximation, the constraint on $w_{\rm pivot}$ exactly matches that which would be obtained in an analysis with constant~$w$.

There is the inconvenience that the pivot scale is dependent on the choice of datasets, though in practice the dependence is not strong \citep{DESIVI,DESIDR2}. It would for instance be a step forward for teams and collaborations to agree a common pivot scale, in the same way the large-scale structure community choose to quote $\sigma_8$, the perturbations on scale $8h^{-1}\,{\rm Mpc}^{-1}$. Pivoting at $a=0.75$  ($z=1/3$), for instance, would be quite suitable for minimizing the unnecessary loss of accuracy when presenting marginalized results -- see for instance \cite{Francis}.

It is well known that no data combination shows significant deviation from $\Lambda$CDM in the case of constant $w$ \citep{DESIVI,DESIDR2}. Hence $w_{\rm pivot}$ is always consistent with $-1$, to an accuracy better than $\pm 0.03$ when all data types are combined, with no significant detection of deviations from $-1$ of the mean equation of state averaged over the observational windows. Hence the evidence against $\Lambda$CDM can be interpreted entirely as a detection significance on the $w_a$ parameter. 

This is quite a remarkable coincidence if $w_a$ is indeed as large as the best-fits suggest. Equivalent statements are that phantom crossing, or the maximum dark energy density ever attained, occur at the centre of the observed window. As we pointed out in \cite{CLDESI}, this is a substantial and unsettling coincidence that is additional to the dark energy coming to dominate at the present epoch.

The visual signature of this coincidence is for the principal axis of the constraint contours to point accurately towards the $\Lambda$CDM point. We are all used to seeing this by now, but need to keep remembering how odd it is. This might induce general skepticism requiring a higher level of proof (see e.g.\ \citealt{CLtension,Wolf,CLDESI,Dinda}), but also prompts investigation of other types of new physics that might address the data, e.g.\ \cite{Craig} and \cite{Chaussidon} amongst many examples.

In any event, when quoting results one ought to decorrelate the amplitude from $w_a$ before marginalising over the latter, to ensure a lossless compression of the constraint. Table V of \cite{DESIDR2} is an example which loses a factor of 2 to 3 in accuracy on the amplitude of $w(a)$ by marginalising onto $w_0$ instead of $w_{\rm pivot}$. By definition, the accuracy on the amplitude in w0waCDM should be the same as in constant wCDM. Without this decorrelation, there is no guarantee that a model that lies within the separate $w_0$ and $w_a$ uncertainties actually fits the data.


\section{Methodologies, or, sigmas from supernovae neither reinforce nor average}

\subsection{Combining exclusion significance over methodological differences}

\cite{DESIDR2} quote three different exclusion limits for $\Lambda$CDM. These combine their own BAO data with the CMB data from {\it Planck} \citep{Planck,Planck2} and ACT \citep{ACT}. and separately with each of the three SNe datasets, Pantheon+ (2.8-sigma exclusion), Union3 (3.8-sigma exclusion), and DESY5 (4.2-sigma exclusion). The results are very similar to the DR1 release, with an extra 0.3-sigma having been added to each significance.

How are we supposed to interpret the overall exclusion, given these three separate results? Tempting options include 
\begin{enumerate}
\item Believe the strongest result.
\item Believe the central result.
\item Average the sigmas.
\end{enumerate}
But none of these have a statistical justification, particularly one that could have been stated before the outcome was known. Our aim is to provide one, which turns out not to support {\em any} of these three options.

When we have datasets that are consistent and independent, the likelihoods multiply, yielding a final constraint which is stronger than the strongest of the individual constraints. But the three SNe datasets have strong overlaps (particularly Pantheon+ and Union3, see Section~\ref{s:supernovae}) and are not independent, as well as having different analysis methodologies.

Now imagine the opposite regime, where the same input data are analysed by three teams under different methodologies, giving consistent but somewhat different results. It’s the same data, so we cannot combine by multiplying. All we can do is choose a probability for each analysis being correct, for example with equal weights. Now the parameter posterior probability distributions {\em add}, not multiply
\begin{equation}
\label{e:weights}
p_{\rm combined}(\bar{\theta}) = \frac{1}{3} p_1(\bar{\theta}) +  \frac{1}{3} p_2(\bar{\theta}) +  \frac{1}{3} p_3(\bar{\theta}) \,,
\end{equation}
where $\bar{\theta}$ is the parameter vector of the model being studied, either the full parameter space or any marginalised subspace such as $w_0$--$w_a$.

In this case, an exclusion result on a fiducial cosmology is dominated by the {\em weakest} result, not the strongest, as it has the largest tail extending beyond the fiducial model. Indeed, in assessing how much probability lies beyond some fiducial point like $\Lambda$CDM, almost all the signal will come from the weakest constraint and is only marginally diminished by the one-third weight.

\begin{figure}
\centering 
\includegraphics[width=6cm]{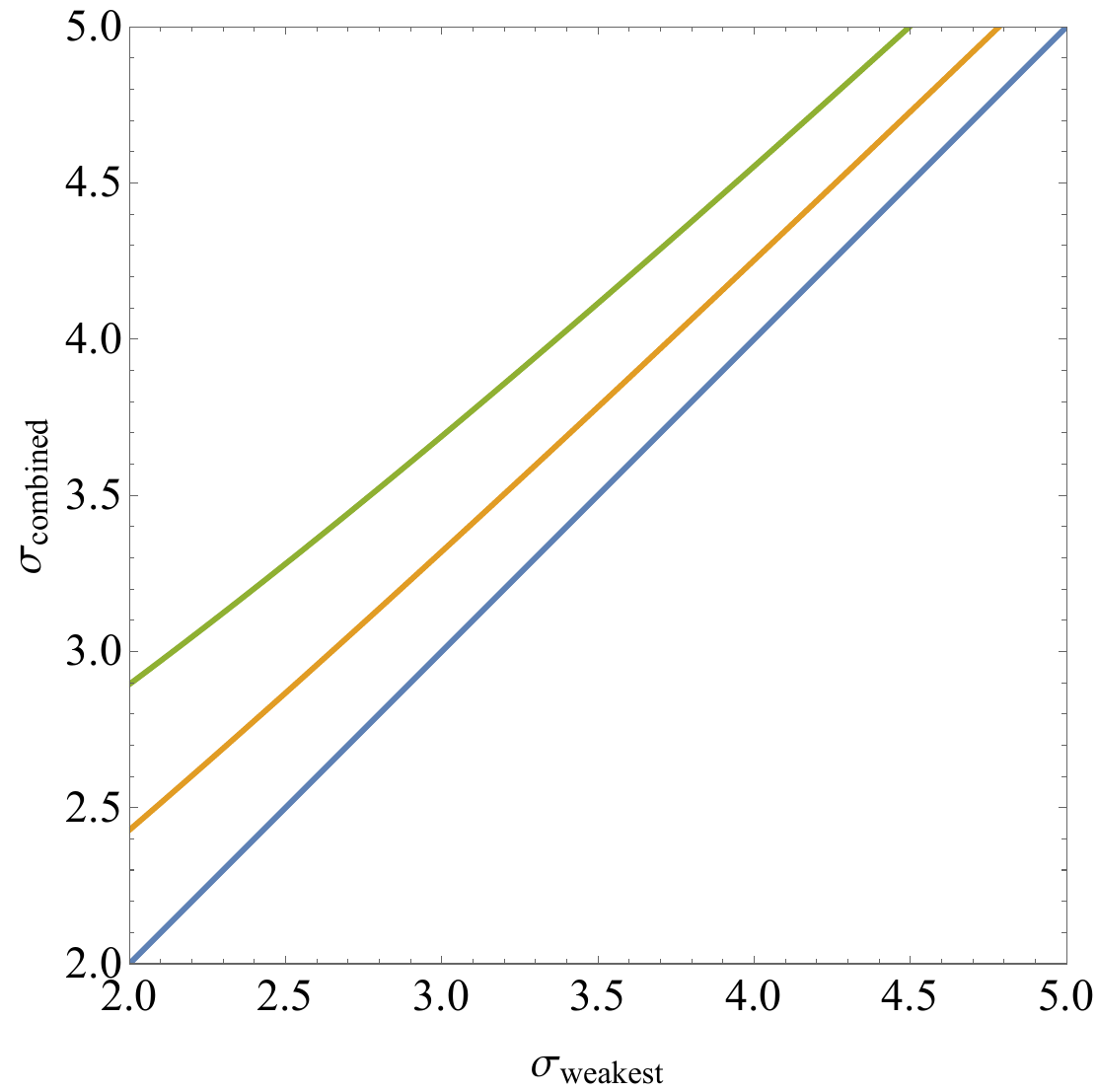}
\caption{\label{f:erfc} The combined significance of methodology-dependent analyses. Blue shows the sigma of the weakest exclusion, orange the combined exclusion for three equally-weighted methodologies, and green if the central outcome is overweighted by a factor of 10. 
}
\end{figure}

Suppose for instance that three such analyses gave 2-sigma, 3-sigma, and 4-sigma exclusions. The first then has 4.6\% probability beyond the fiducial point and the other two much less. In the weighted combined probability there is still 1.5\% remaining beyond the fiducial point, and this corresponds only to 2.4-sigma. The strength of the 3-sigma and 4-sigma results is unimportant. In the conservative limit where we assume the stronger analyses contribute negligibly, the combined significance can be written using the complementary error function as\footnote{This assumes gaussianity; if there happen also to be fat tails to the distribution the significance will be further reduced.}
\begin{equation}
\label{e:erfcequal}
\sigma_{\rm combined} = \sqrt{2} \,{\rm Erfc}^{-1} \left[ \frac{1}{3} {\rm Erfc} \left[ \frac{\sigma_{\rm weakest}}{\sqrt{2}} \right] \right] \,,
\end{equation}
which is plotted in Figure~\ref{f:erfc}. The combined sigma, in orange, is hardly greater than the weakest sigma, in blue. The result is the same whether one-sided or two-sided exclusions are used.



The above discussion is framed for a fiducial model embedded in a one-parameter extension, where one asks what fraction of the probability distribution extends beyond the fiducial point. DESI are dealing with the two-parameter $w_0$--$w_a$ extension, and assess significance using the $\Delta \chi^2_{\rm MAP}$ statistic that asks how much worse is the (marginalised) $\Lambda$CDM fit as compared to the best-fit \citep{DESIDR2}. However they then express this as the 1D equivalent in sigmas (their equation 22), which ensures the above argument applies as equation (\ref{e:erfcequal}) holds in any subspace of the full model.

One might argue that the weights should not all be taken equal, for example that the middle result is the most credible. But even if we permitted this dubious {\it a posteriori} reasoning and gave it ten times the probability of the others, we would still only have 
\begin{equation}
\sigma_{\rm combined} = \sqrt{2} \, {\rm Erfc}^{-1} \left[ \frac{1}{12} {\rm Erfc}\left[\frac{\sigma_{\rm weakest}}{\sqrt{2}} \right] \right] \,,
\end{equation}
also shown in Figure~\ref{f:erfc}. This doesn’t make much difference.

The conclusion is that when multiple analyses of the same dataset give varying outcomes, the inferred exclusion significance, in sigmas, should be guided by the weakest result, not an average or the best result. The effect becomes more extreme as the confidence level increases.

The weighted combination of the posteriors seen in equation (\ref{e:weights}) bears some resemblence to Bayesian Model Averaging (see \cite{PL13} for a review), but there the dataset is fixed and different models featuring common parameters are combined to give a single posterior on a shared parameter. In that case posterior model weights are used in the averaging, but that is not possible here as there is no way to assign posterior probabilities to each dataset choice. Another related technique is hyperparameters \citep{Hobson}, which can be used to set a probability that a given dataset within a compilation of independent data is correct given specific model assumptions, but that again is not available here because the datasets in question are not independent.

\subsection{Implications for DESI DR2}

So what does this mean for the DESI DR2 analysis? The use of SNe datasets in the DESI DR2 analysis lies somewhere between data independence and methodological differences.  A detailed expert analysis may be able to unpick these distinctions, but this has not yet been done. A useful first step would be to compare the different methodologies on a single agreed dataset. 

The interpretation is also made difficult because the final outcome is a marginalisation after complicated intersections of preferred regions in the full parameter space. These are not well represented by what is going on in the marginalised $w_0$--$w_a$ space. Moreover SNe data on their own are so unconstraining in the $w_0$--$w_a$ plane that, of the three, only DESY5 \citep{DESY5} even shows a plot of these constraints from the SNe alone (Pantheon+ does tabulate a result combined with SH0ES).

The quoted exclusions of $\Lambda$CDM for DESI+CMB+SNe are 2.8-sigma (Pantheon+), 3.8-sigma (Union3), and 4.2-sigma (DESY5). The dataset overlap is severe, because not only do the SNe samples overlap, but the combinations have perfect overlap of the DESI and CMB datasets. Multiplying the likelihoods is therefore not an option. Hence the combined probability must be obtained through addition as above, where equal weights appear appropriate given that the DESI paper handles the three SNe samples even-handedly. Doing so, the statistical exclusion of $\Lambda$CDM from equation~(\ref{e:erfcequal}) is 3.1-sigma (dominated by Pantheon+, the other two datasets serving only to contribute the weight factors). 

It is tempting to instead average the three quoted sigmas, or take the middle one as indicative. But this is not what principled statistical inference implies. Coincidentally (or perhaps not entirely so), 3.1-sigma is also the quoted exclusion for DESI+CMB {\em without} any supernova data.\footnote{Adding the supernovae does significantly tighten the overall constraint in the \mbox{$w_0$--$w_a$} plane, principally by cutting off the regions furthest from $\Lambda$CDM. It is the exclusion level of $\Lambda$CDM that does not change.}

We acknowledge that this is conservative, as the three SNe datasets have only partial overlap. Hence their combination should yield some additional statistical power. But this has to be quantitatively demonstrated before it can be claimed, with the systematic uncertainty from methodology differences properly accounted for. In the next section we make some comments on the current obstacles to doing so.

Can't we pick one SNe dataset, such as DESY5 with its impressive 4.2-sigma? No, no more than you can expect a Las Vegas casino to let you spin the roulette wheel three times before deciding which ball determines your bet! It would have been perfectly legitimate to decide in advance of analysis that one dataset and its methodology are superior, and analyse only that one. But it is not statistically acceptable to select one's favourite dataset only after the analysis of all three has taken place.

To summarise again, when datasets are independent and consistent, one gets an exclusion result somewhat stronger than the strongest individual result. When a single dataset is analysed with different methodologies, the combined result is somewhat stronger only than the {\em weakest} result. While the DESI analysis lies between these two limits, we cannot combine as independent and so only the second, more conservative, approach is available. Based on this reasoning, the best estimate of the exclusion significance on $\Lambda$CDM from the DESI DR2 analysis is 3.1-sigma.

\section{The supernovae compilations}
\label{s:supernovae}

Having reached our conclusion on overall exclusion significance, we now study the samples for trends and obstacles to future combination. It's time for a closer look.

The three alternate SNe compilations used by DESI are of comparable size but with quite distinct properties:
\begin{description} 
\item[{\bf Pantheon+}] \citep{PantheonPlus,Brout22} consists of 1550 unique spectroscopically-identified supernovae drawn from many surveys and analysed using a version of the SALT2 light-curve fitter \citep{Guy}. 
\item[{\bf Union3}] \citep{Union3} is also spectroscopic. It is larger, with 2087 unique supernovae including most (1363 members) of the Pantheon+ compilation. It uses the SALT3 light-curve fitter \citep{SALT3,Taylor} and a quite different hierarchical Bayesian analysis scheme UNITY \citep{UNITY}. 
\item[{\bf DESY5}] \citep{DESY5,DESY5data} uses photometric selection, with 1635 supernovae of their own acquisition that are all (with three exceptions) at $z>0.1$. In order to anchor the Hubble diagram at low redshift, a separate set of 194 local supernovae from historical samples is added, almost all of which are also in the other two compilations. The analysis methodology uses SALT3 but is otherwise similar to that of Pantheon+. A small fraction of the new DES SNe are also included in the other two compilations via separate spectroscopic observations.
\end{description}

\subsection{DESY5, high- and low-redshift}

Some pause for thought is given by the central panel of Figure 14 of \cite{DESIDR2}. While the full DESY5 compilation leads to the strongest overall exclusion, when the sample is restricted to the $z>0.1$ sample actually obtained (photometrically) by DES, the exclusion significance falls to only just over 2-sigma. It is the low-redshift addition to the sample, in strong overlap with the other compilations, which is responsible for the high exclusion result. 

\begin{figure}
\centering 
\includegraphics[width=7cm]{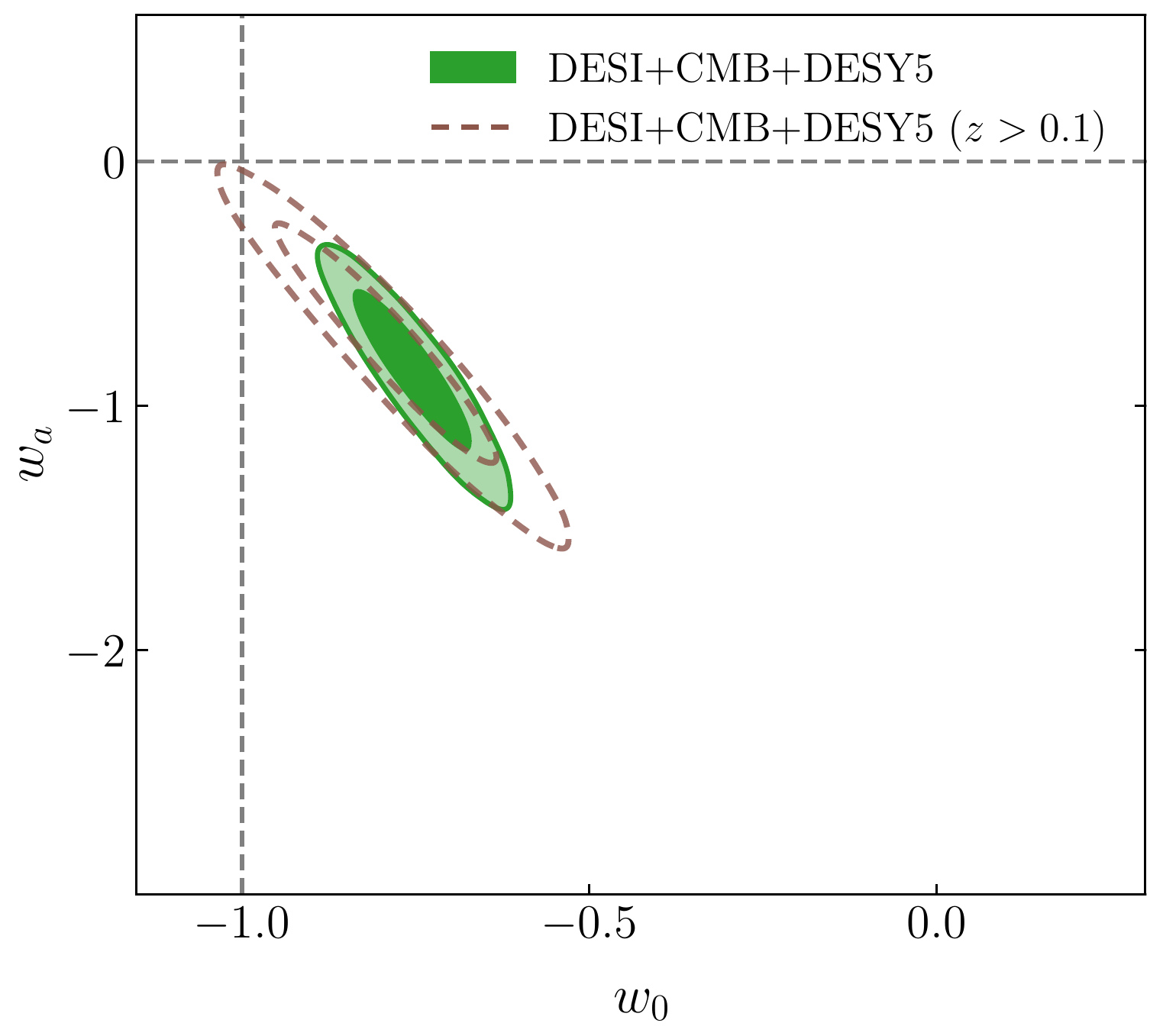}
\caption{\label{f:DESY5} Constraint contours for DESI+CMB+DESY5, showing the full sample and with the low-redshift sample excised. [Adapted from Figure~14 of \protect\cite{DESIDR2} under Creative Commons BY 4.0 License, original caption ``The change to the DESI+CMB+DESY5 posterior (green) when excluding low-redshift SNe at $z<0.1$ from the DESY5 sample (brown dashed). The uncertainties are much larger when excluding these SNe but the shift in the best-fit values of $w_0$ and $w_a$ is small.'']
}
\end{figure}

We reproduce their figure here as Figure~\ref{f:DESY5} to emphasise this. There seems to be some level of inadvertent denial at play in the DESI paper's description of this, as it focuses on the irrelevant point that the best-fit $w_0$--$w_a$ is almost unchanged rather than the crucial point that the exclusion significance is much reduced, and that the DESY5 SNe alone are pulling DESI+CMB {\em towards} $\Lambda$CDM, not away (as to a lesser extent does Pantheon+). This was already noted for the DESI DR1 analysis by \cite{Gialamas} and subsequent papers discussed below. Figure 13 of \cite{DESIDR2} shows that this situation remains.

Prompted by \cite{EfstSN}'s uncovering of a possible calibration mismatch in the low-redshift common DESY5/Pantheon+ sample (see also \cite{Huang} and our next subsection), \cite{Notari} constructed a `split' DESY5 likelihood that allows the relative calibration of the low- and high-redshift subsamples to float. This essentially removes the anchoring effect of the low-redshift sample and weakens the exclusion (to 1.7-sigma for an analysis using the DESI DR1 BAO data, after marginalising over the relative offset). We wish to empasise here that this mimics removing the low-redshift sample entirely. 

\cite{Vincenzi} re-assert their confidence in the DESY5 calibration in light of the above papers, through tracking of analysis improvements and differences in selection. But even if so, this still leaves us in a position where the actual DES SNe alone are pulling DESI+CMB quite strongly towards $\Lambda$CDM, with the low-redshift sample being largely responsible for the highest quoted exclusion.

\subsection{Pantheon+ versus DESY5}

Given their relatively small overlap and comparable methodologies, Pantheon+ and DESY5 are the best test case for a combined study. This has been attempted by \cite{Notari}, in the context of the DESI DR1 results. This study already gives strong indications of stumbling blocks on the route to a full joint analysis. They sought to combine those two datasets by eliminating the overlap of 335 SNe from one or other dataset, then carrying out an analysis which included both.

However, the ultimate result is quite different depending on whether the overlapping SNe are excised from the DESY5 sample or from Pantheon+ \citep{Notari}. If the reduced DESY5 dataset is added to Pantheon+, the exclusion remains at 2.5-sigma as for Pantheon+ alone (remember this is for DESI DR1). If instead the reduced Pantheon+ dataset is added to DESY5, the exclusion of 3.8-sigma almost matches the 3.9-sigma from DESY5 alone. In each case the addition of the extra data does improve the overall constraining power, but the best-fit also shifts in such a way that the exclusion of $\Lambda$CDM is practically unchanged.

This is quite striking. The incorporated datasets are now identical, yet simply switching 10\% of the SNe between the two methodologies changes the exclusion by more than 1-sigma. This is despite the entire DESI+CMB datasets {\em and} 90\% of the SNe being treated identically. Further, these are the two surveys whose methodologies are advertised as the most similar, though they do use different versions of the SALT light-curve fitter. \cite{Vincenzi} however caution that altering the samples will change the selection criteria and requires full recomputation of bias corrections. It is clear that robust combination of the existing SNe datasets is a major undertaking.

\subsection{Pantheon+ versus Union3}

In the Union3 analysis of \cite{Union3}, they note that they end up with higher uncertainties than Pantheon+, despite having a larger sample which contains nearly all of Pantheon+. They point to specific additional uncertainties they include which are responsible for this, most importantly intergalactic extinction and the redshift evolution of correlations with host-galaxy properties. This creates a case that the Pantheon+ uncertainties may have some systematic under-estimation.\footnote{Separately, \cite{Kim} has noted a difference between DESI's priors and those used to derive the Union3 spline-interpolated posterior distribution in the DESI DR1 release, but did not find significant impact.}

Ironically though, since Pantheon+ is pulling DESI+CMB mildly {\em towards} $\Lambda$CDM, increasing their uncertainties is likely to have the effect of {\em increasing} the combined exclusion of $\Lambda$CDM towards the 3.1-sigma of DESI+CMB alone.

\section{Conclusion}

The current stance facing the cosmology community on the combination of SNe data with other probes is complex. We use DESI DR2 as a case study, with the goal of showing the effect of combining state-of-the-art SNe datsets with other cosmology probes. We provide a principle-based argument which indicates that, for DESI DR2, the best inferred exclusion of $\Lambda$CDM taking the three SNe datasets into account is 3.1-sigma. Pantheon+ and the actual DESY5 sample (excluding the low-redshift historical additions), both in fact pull the DESI+CMB result towards $\Lambda$CDM.

It has been established beyond doubt that the Supernovae played a decisive role in establishing the current accelerated expansion of the universe. \citep{Riess,Perlmutter}. Today, they continue to provide important cosmological constraints despite having entered a systematics-dominated regime with all the modeling challenges that that entails. Yet new challenges emerge if SNe are to remain important in the forthcoming era of Stage 5 surveys \citep{Stage5}. Users of the SNe data have to deal with competing dataset compilations and assess choices of modeling methodologies. A comparison of the different methodologies on a standardised agreed dataset would be extremely useful. Establishing a more secure low-redshift anchor would also help eliminate dependence on historical samples with uneven selection properties and characteristics.

We conclude that while we can quote a best-estimate exclusion of 3.1-sigma including the SNe data, the most secure inference from the analysis is the same 3.1-sigma exclusion level obtained from DESI+CMB alone.

\section*{Acknowledgements}
We thank Ed Copeland, George Efstathiou, Alex Kim, and Martin White for discussions.  This work was supported by the Funda\c{c}\~{a}o para a Ci\^encia e a Tecnologia (FCT) through the research grants UIDB/04434/2020 and UIDP/04434/2020. M.C.\ acknowledges support from the FCT through the Investigador FCT Contract No.\ CEECIND/02581/2018 and POPH/FSE (EC). A.R.L.\ acknowledges support from the FCT through the Investigador FCT Contract No.\ CEECIND/02854/2017 and POPH/FSE (EC). This article/publication is based upon work from COST Action CA21136 -- ``Addressing observational tensions in cosmology with systematics and fundamental physics (CosmoVerse)'', supported by COST (European Cooperation in Science and Technology). This work benefited from the EuCAPT rapid response workshop on April 14$^{\rm th}$ 2025; particular thanks to Kyle Dawson, Colin Hill, and Eiichiro Komatsu.
\section*{Data Availability}

No new data were generated or analysed in support of this research.




\begin{thebibliography}{}
\bibitem[\protect\citeauthoryear{Abbott et al.}{2024}]{DESY5} Abbott T. M. C. et al. (DES collaboration), 2024, ApJL 973, L14, \emph{The Dark Energy Survey: Cosmology Results With $\sim$1500 New High-redshift Type Ia Supernovae Using The Full 5-year Dataset}, \href{https://arXiv.org/abs/2401.02929}{arXiv:2401.02929 [astro-ph.CO]}
\bibitem[\protect\citeauthoryear{Abdul-Karim et al.}{2025}]{DESIDR2} Abdul-Karim M. et al. (DESI Collaboration), 2025, Phys. Rev. D 112, 083515, \emph{DESI DR2 Results II: Measurements of Baryon Acoustic Oscillations and Cosmological Constraints}, \href{https://arxiv.org/abs/2503.14738}{	arXiv:2503.14738 [astro-ph.CO]}
\bibitem[\protect\citeauthoryear{Adame et al.}{2025}]{DESIVI} Adame A. G. et  al. (DESI Collaboration), 2025, JCAP 02, 021, \emph{DESI 2024 VI: Cosmological Constraints from the Measurements of Baryon Acoustic Oscillations}, \href{https://arXiv.org/abs/2404.03002}{arXiv:2404.03002 [astro-ph.CO]} 
\bibitem[\protect\citeauthoryear{Aghanim et al.}{2020}]{Planck} Aghanim N. et al. (Planck Collaboration), 2020, A\&A 641, A6, \emph{Planck 2018 results. VI. Cosmological parameters}, \href{https://arXiv.org/abs/1807.06209}{arXiv:1807.06209 [astro-ph.CO]}
\bibitem[\protect\citeauthoryear{Akrami et al.}{2020}]{Planck2} Akrami Y. et al. (Planck Collaboration), 2020, A\&A 643, A42, \emph{Planck intermediate results. LVII. Joint Planck LFI and HFI data processing}, \href{https://arXiv.org/abs/2007.04997}{[arXiv:2007.04997 [astro-ph.CO]]}
\bibitem[\protect\citeauthoryear{Albrecht et al.}{2006}]{DETF} Albrecht A. et al., 2006, \emph{Report of the Dark Energy Task Force}, \href{https://arXiv.org/abs/astro-ph/0609591}{arXiv:astro-ph/0609591}
\bibitem[\protect\citeauthoryear{Besuner et al.}{2025}]{Stage5} Besuner R. et al., 2025, \emph{The Spectroscopic Stage-5 Experiment}, \href{https://arxiv.org/abs/2503.07923}{	arXiv:2503.07923 [astro-ph.CO]}
\bibitem[\protect\citeauthoryear{Brout et al.}{2022}]{Brout22} Brout D. et al. (Pantheon+ collaboration), 2022, ApJ 938, 110, \emph{The Pantheon+ Analysis: Cosmological Constraints}, \href{https://arxiv.org/abs/2202.04077}{arXiv:2202.04077 [astro-ph.CO]}
\bibitem[\protect\citeauthoryear{Calderon et al.}{2024}]{CalderonDR1} Calderon R. et al., 2024, JCAP 10, 048, \emph{DESI 2024: Reconstructing Dark Energy using Crossing Statistics with DESI DR1 BAO data}, \href{https://arxiv.org/abs/2405.04216}{arXiv:2405.04216 [astro-ph.CO]}
\bibitem[\protect\citeauthoryear{Caldwell}{2002}]{Caldwell} Caldwell R. R., 2002, Phys. Lett. B545, 23, \emph{A Phantom Menace? Cosmological consequences of a dark energy component with super-negative equation of state}, \href{https://arxiv.org/abs/astro-ph/9908168}{arXiv:astro-ph/9908168}
\bibitem[\protect\citeauthoryear{Chaussidon et al.}{2025}]{Chaussidon} Chaussidon E. et al., 2025, \emph{Early time solution as an alternative to the late time evolving dark energy with DESI DR2 BAO}, \href{https://arxiv.org/abs/2503.24343}{	arXiv:2503.24343 [astro-ph.CO]}
\bibitem[\protect\citeauthoryear{Chevallier \& Polarski}{2001}]{CPL1} Chevallier M., Polarski D., 2001, Int. J. Mod. Phys. D10, 213, \emph{Accelerating Universes with Scaling Dark Matter}, \href{https://arXiv.org/abs/gr-qc/0009008}{arXiv:gr-qc/0009008}
\bibitem[\protect\citeauthoryear{Cort\^es \& Liddle}{2024a}]{CLtension} Cort\^es M, Liddle A. R., 2024a, MNRAS 531, L52, \emph{On dataset tensions and signatures of new cosmological physics}, \href{https://arxiv.org/abs/2309.03286}{	arXiv:2309.03286 [astro-ph.CO]}
\bibitem[\protect\citeauthoryear{Cort\^es \& Liddle}{2024b}]{CLDESI} Cort\^es M, Liddle A. R., 2024b, JCAP 12, 007, \emph{Interpreting DESI's evidence for evolving dark energy}, \href{https://arxiv.org/abs/2404.08056}{	arXiv:2404.08056 [astro-ph.CO]}
\bibitem[\protect\citeauthoryear{Craig et al.}{2024}]{Craig} Craig N., Green D., Meyers J., Rajendran S., 2024, JHEP 09, 097, \emph{No $\nu$s is Good News}, \href{https://arxiv.org/abs/2405.00836}{arXiv:2405.00836 [astro-ph.CO]}
\bibitem[\protect\citeauthoryear{Dinda et al.}{2025}]{Dinda} Dinda B. R., Maartens R., Saito S., Clarkson C., 2025, 	JCAP 08, 018, \emph{Improved null tests of $\Lambda$CDM and FLRW in light of DESI DR2}, \href{https://arxiv.org/abs/2504.09681}{arXiv:2504.09681 [astro-ph.CO]}
\bibitem[\protect\citeauthoryear{Efstathiou}{2025}]{EfstSN} Efstathiou G., 2025, MNRAS 538, 875, \emph{Evolving Dark Energy or Supernovae Systematics?}, \href{https://arxiv.org/abs/2408.07175}{	arXiv:2408.07175 [astro-ph.CO]}
\bibitem[\protect\citeauthoryear{Francis et al.}{2007}]{Francis} Francis M. J., Lewis G. F., Linder E. V., 2007, MNRAS 380, 1079,  \emph{Power Spectra to 1\% Accuracy between Dynamical Dark Energy Cosmologies}, \href{https://arxiv.org/abs/0704.0312}{arXiv:0704.0312 [astro-ph]}
\bibitem[\protect\citeauthoryear{Gan et al.}{2025}]{Gan} Gan G. et al. (DESI Collaboration), 2025, Nature Astronomy, \emph{Dynamical Dark Energy in light of the DESI DR2 Baryonic Acoustic Oscillations Measurements}, \href{https://arxiv.org/abs/2504.06118}{arXiv:2504.06118 [astro-ph.CO]}
\bibitem[\protect\citeauthoryear{Gialamas et al.}{2024}]{Gialamas} Gialamas I. D., H\"utsi G., Kannike K., Racioppi A., Raidal M., Vasar M., Veermäe H., 2025, Phys. Rev. D 111, 043540, \emph{Interpreting DESI 2024 BAO: Late-time dynamical dark energy or a local effect?}, \href{https://arxiv.org/abs/2406.07533}{	arXiv:2406.07533 [astro-ph.CO]}
\bibitem[\protect\citeauthoryear{Guy et al.}{2007}]{Guy}  Guy J. et al., 2007, A\&A, 466, 11, \emph{SALT2: using distant supernovae to improve the use of Type Ia supernovae as distance indicators}, \href{https://arXiv.org/abs/astro-ph/0701828}{arXiv:astro-ph/0701828}
\bibitem[\protect\citeauthoryear{Hobson et al.}{2002}]{Hobson} Hobson M. P., Bridle S. L., Lahav O, 2002, MNRAS, 335, 377, \emph{Combining cosmological datasets: hyperparameters and Bayesian evidence}, \href{https://arxiv.org/abs/astro-ph/0203259}{arXiv:astro-ph/0203259}
\bibitem[\protect\citeauthoryear{Huang et al.}{2025}]{Huang} Huang L., Cai R.-G., Wang S.-J, 2025, Sci. China-Phys. Mech. Astron. 68, 100413, \emph{The DESI 2024 hint for dynamical dark energy is biased by low-redshift supernovae}, \href{https://arxiv.org/abs/2502.04212}{arXiv:2502.04212 [astro-ph.CO]}
\bibitem[\protect\citeauthoryear{Huterer \& Turner}{2001}]{HT} Huterer D., Turner M. S., 2001,  Phys. Rev. D64, 123527, \emph{Probing the dark energy: methods and strategies}, \href{https://arXiv.org/abs/astro-ph/0012510}{arXiv:astro-ph/0012510}
\bibitem[\protect\citeauthoryear{Kenworthy et al.}{2021}]{SALT3} Kenworthy W. D. et al., 2021, ApJ 923, 265, \emph{SALT3: An Improved Type Ia Supernova Model for Measuring Cosmic Distances}, \href{https://arxiv.org/abs/2104.07795}{	arXiv:2104.07795 [astro-ph.CO]}
\bibitem[\protect\citeauthoryear{Kim}{2024}]{Kim} Kim A. G., 2024, \emph{Comments on the Union3 ``Spline-Interpolated Distance Moduli'' Model}, \href{https://arxiv.org/abs/2412.14181}{	arXiv:2412.14181 [astro-ph.CO]}
\bibitem[\protect\citeauthoryear{Linder}{2003}]{CPL2} Linder E. V., 2003, Phys. Rev. Lett. 90, 091301, \emph{Exploring the Expansion History of the Universe}, \href{https://arXiv.org/abs/astro-ph/0208512}{arXiv:astro-ph/0208512}
\bibitem[\protect\citeauthoryear{Linder}{2006}]{LinderPivot} Linder E. V., 2006, Astropart. Phys. 26, 102,  \emph{Biased Cosmology: Pivots, Parameters, and Figures of Merit}, \href{https://arXiv.org/abs/astro-ph/0604280}{arXiv:astro-ph/0604280}
\bibitem[\protect\citeauthoryear{Lodha et al.}{2025a}]{LodhaDR1} Lodha K. et al. (DESI Collaboration), 2025a, Phys. Rev. D 111, 023532, \emph{DESI 2024: Constraints on Physics-Focused Aspects of Dark Energy using DESI DR1 BAO Data}, \href{https://arxiv.org/abs/2405.13588}{	arXiv:2405.13588 [astro-ph.CO]}
\bibitem[\protect\citeauthoryear{Lodha et al.}{2025b}]{Lodha} Lodha K. et al., 2025b, \emph{Extended Dark Energy analysis using DESI DR2 BAO measurements}, \href{https://arxiv.org/abs/2503.14743}{arXiv:2503.14743 [astro-ph.CO]}
\bibitem[\protect\citeauthoryear{Madhavacheril et al.}{2024}]{ACT} Madhavacheril M. S. et al. (ACT Collaboration), 2024, ApJ 962, 113, \emph{The Atacama Cosmology Telescope: DR6 Gravitational Lensing Map and Cosmological Parameters}, \href{https://arxiv.org/abs/2304.05203}{arXiv:2304.05203 [astro-ph.CO]}
\bibitem[\protect\citeauthoryear{Martin \& Albrecht}{2006}]{MartinPivot} Martin D., Albrecht A., 2006, \emph{Talk about Pivots}, \href{https://arXiv.org/abs/astro-ph/0604401}{arXiv:astro-ph/0604401}
\bibitem[\protect\citeauthoryear{Notari et al.}{2025}]{Notari} Notari A., Redi M., Tesi A., 2025, JCAP 04, 048, \emph{BAO vs. SN evidence for evolving dark energy}, \href{https://arxiv.org/abs/2411.11685}{	arXiv:2411.11685 [astro-ph.CO]}
\bibitem[\protect\citeauthoryear{Parkinson \& Liddle}{2013}]{PL13} Parkinson D., Liddle A. R., 2013, Statistical Analysis and Data Mining 6, 3, \emph{Bayesian Model Averaging in Astrophysics: A Review}, \href{https://arxiv.org/abs/1302.1721}{	arXiv:1302.1721 [astro-ph.IM]}.
\bibitem[\protect\citeauthoryear{Perlmutter et al.}{1999}]{Perlmutter} Perlmutter S. et al. (The Supernova Cosmology Project), 1999, ApJ, 517, 565, \emph{Measurements of Omega and Lambda from 42 High-Redshift Supernovae}, \href{https://arxiv.org/abs/astro-ph/9812133}{arXiv:astro-ph/9812133}
\bibitem[\protect\citeauthoryear{Riess et al.}{1998}]{Riess} Riess A. et al., 1998, AJ, 116, 1009, \emph{Observational Evidence from Supernovae for an Accelerating Universe and a Cosmological Constant}, \href{https://arxiv.org/abs/astro-ph/9805201}{arXiv:astro-ph/9805201}
\bibitem[\protect\citeauthoryear{Rubin et al.}{2015}]{UNITY} Rubin D. et al., 2015, ApJ 813, 137, \emph{UNITY: Confronting Supernova Cosmology's Statistical and Systematic Uncertainties in a Unified Bayesian Framework}, \href{https://arxiv.org/abs/1507.01602}{arXiv:1507.01602 [astro-ph.CO]}
\bibitem[\protect\citeauthoryear{Rubin et al.}{2023}]{Union3} Rubin D. et al., 2023, \emph{Union Through UNITY: Cosmology with 2,000 SNe Using a Unified Bayesian Framework}, \href{https://arXiv.org/abs/2311.12098}{arXiv:2311.12098 [astro-ph.CO]} 
\bibitem[\protect\citeauthoryear{S\'anchez et al.}{2024}]{DESY5data} S\'anchez B. O. et al. (DES collaboration), 2024, ApJ 975, 5, \emph{The Dark Energy Survey Supernova Program: Light curves and 5-Year data release}, \href{https://arxiv.org/abs/2406.05046}{arXiv:2406.05046 [astro-ph.CO]}
\bibitem[\protect\citeauthoryear{Scolnic et al.}{2022}]{PantheonPlus} Scolnic D. et al., 2022, ApJ 938, 113, \emph{The Pantheon+ Analysis: The Full Data Set and Light-curve Release},  \href{https://arXiv.org/abs/2112.03863}{arXiv:2112.03863 [astro-ph.CO]}
\bibitem[\protect\citeauthoryear{Taylor et al.}{2023}]{Taylor} Taylor G. et al., 2023, MNRAS 520, 5209, \emph{SALT2 versus SALT3: updated model surfaces and their impacts on type Ia supernova cosmology}, \href{https://arxiv.org/abs/2301.10644}{	arXiv:2301.10644 [astro-ph.CO]}
\bibitem[\protect\citeauthoryear{Vincenzi et al.}{2025}]{Vincenzi} Vincenzi M. et al. (DES Collaboration), 2025, \emph{Comparing the DES-SN5YR and Pantheon+ SN cosmology analyses: Investigation based on ``Evolving Dark Energy or Supernovae systematics?''}, \href{https://arxiv.org/abs/2501.06664}{	arXiv:2501.06664 [astro-ph.CO]}
\bibitem[\protect\citeauthoryear{Wolf \& Ferreira}{2023}]{Wolf} Wolf W., Ferreira P. G., 2023, Phys. Rev. D 108, 103519, \emph{Underdetermination of dark energy}, \href{https://arxiv.org/abs/2310.07482}{arXiv:2310.07482 [astro-ph.CO]}

\bsp	
\label{lastpage}
\end{thebibliography}


\end{document}